\documentclass[aip,apl,amsmath,amssymb,reprint,]{revtex4-1}
\usepackage{graphicx,psfrag}
\usepackage{dcolumn}
\usepackage{bm}
\usepackage{color}

\begin{document}

\preprint{AIP/123-QED}

\title{Impact of second-order piezoelectricity on electronic and optical properties of $c$-plane In$_{x}$Ga$_{1-x}$N quantum dots: Consequences for long wavelength emitters}

\author{Saroj Kanta Patra}
\email{sarojkanta.patra@tyndall.ie}
 \affiliation {Department of Electrical Engineering, University College Cork}
 \affiliation{
Tyndall National Institute, Lee Maltings, Dyke Parade, Cork
}%
\author{Stefan Schulz}%
 \affiliation{
Tyndall National Institute, Lee Maltings, Dyke Parade, Cork
}%

\date{\today}

\begin{abstract}
In this work we present a detailed analysis of the second-order
piezoelectric effect in $c$-plane In$_{x}$Ga$_{1-x}$N/GaN quantum
dots and its consequences for electronic and optical properties of
these systems. Special attention is paid to the impact of increasing
In content $x$ on the results. We find that in general the
second-order piezoelectric effect leads to an increase of the
electrostatic built-in field. Furthermore, our results show that for
an In content $\geq$30\% this increase in the built-in field has a
significant effect on the emission wavelength and the radiative
lifetimes. For instance, at 40\% In, the radiative lifetime is more
than doubled when taking second-order piezoelectricity into account.
Overall our calculations reveal that when designing and describing
the electronic and optical properties of $c$-plane
In$_{x}$Ga$_{1-x}$N/GaN quantum dot based light emitters with high
In contents, second-order piezoelectric effects cannot be neglected.
%
\end{abstract}

\maketitle

Heterostructures based on In$_{x}$Ga$_{1-x}$N alloys are of great
technological interest thanks to its bandgap energy $E_g$ tunability
with changing In content $x$.~\cite{NaSe1996,MaNe2011,NaFa1997} By
changing $x$, in principle, wavelengths between ultraviolet
($E^\text{GaN}_g\approx3.5$ eV) and the near infrared regime
($E^\text{InN}_g\approx0.67$ eV) are achievable. Thus, this ideally
broad range spans the entire visible spectrum, rendering the
material highly suitable for use in optoelectronic devices such as
light-emitting diodes, laser diodes and solar
cells.~\cite{NaSe1996,MaNe2011,NaFa1997} Utilizing
In$_{x}$Ga$_{1-x}$N/GaN based quantum well (QW) structures, high
efficiency devices operating in the violet and blue spectral region
have been realized over the last few years.~\cite{Na2015,ShCa2015}
However, keeping the efficiency high and extending the emission
wavelength into the green, yellow or infrared wavelength range by
increasing the In content $x$ is still
challenging.~\cite{JiLi2015,DuMa2014} Several factors contribute to
this so-called ``green gap'' problem,~\cite{HuGr2017} ranging from
sample quality, due to large strains in QW systems, up to strong
strain-induced electrostatic piezoelectric fields. The piezoelectric
effect leads, for instance, to a spatial separation of electron and
hole wave functions and in turn to increased radiative lifetimes,
also known as the quantum confined Stark effect
(QCSE).~\cite{WaXi2001} However, theoretical studies have shown that
the built-in field is strongly reduced in an In$_{x}$Ga$_{1-x}$N/GaN
quantum dot (QD) when compared to a QW of the same height and In
content.~\cite{ScRe2010} This originates from strain relaxation
mechanism and surface area effects, stemming from the three
dimensional QD confinement, and results in a reduction of the QCSE
in QDs. Therefore, for an In$_{x}$Ga$_{1-x}$N/GaN QD, when compared
to a In$_{x}$Ga$_{1-x}$N/GaN QW of the same height, the In content
$x$ in the dot can be increased considerably for a comparable field
in both systems. This suggests that In$_{x}$Ga$_{1-x}$N/GaN QDs are
promising candidates to achieve efficient radiative recombination at
longer wavelength. Recently, making use of this concept,
In$_{x}$Ga$_{1-x}$N/GaN QD based light emitters operating in the
green to yellow spectral range have been
realized.~\cite{WeMe2016,MeWe2017} Moreover, Frost and
co-workers~\cite{FrHa2016,SuFr2015} have demonstrated high
performance red emitting ($>$630 nm) lasers using
In$_{x}$Ga$_{1-x}$N/GaN QDs. To achieve this emission wavelength, In
contents as high as 40\% have been reported.~\cite{FrHa2016} Only a
few theoretical studies have addressed the electronic and optical
properties of these high In content, long wavelength
emitters.~\cite{SuFr2015,KhSo2010} Additionally, piezoelectric
fields in thin InN layers, embedded in GaN, have been used to
achieve topological insulator states.~\cite{MiYa2012} However, all
previous theoretical studies on these different aspects of InGaN/GaN
based systems included linear piezoelectric polarization
contributions only. Recently, non-linear piezoelectricity effects
have been discussed and reported for wurtzite III-N QW
systems.~\cite{PaTs2011,PrWa2013,MiPa2014} These studies showed for
instance better agreement between theoretical and experimental
built-in field values when second-order piezoelectric effects are
considered in the calculations. Furthermore, in other material
systems, such as zincblende InAs/GaAs QDs, second-order
piezoelectric effects have been highlighted to affect their
electronic and optical properties
significantly.~\cite{BeZu2006,BeWu2006,ScWi2007} However, no
detailed study exists on the importance of second-order
piezoelectricity on electronic and optical properties of $c$-plane
In$_{x}$Ga$_{1-x}$N dots with varying In content. Given the recent
drive for In$_{x}$Ga$_{1-x}$N QD based light emitters with high In
contents ($x\approx0.4$), the question of how important second-order
piezoelectric effects are for describing and designing these
emitters for future optoelectronic devices is of central importance.

Here, we address this question by calculating electronic and optical
properties of $c$-plane In$_{x}$Ga$_{1-x}$N/GaN QDs with In contents
ranging from 10\% to 50\% using a continuum based model, including
the full second-order piezoelectric polarization vector. We find
that the second-order piezoelectric effect leads to an increase in
the electrostatic built-in field when compared to the situation
where only standard first-order contributions are accounted for.
This increase in the built-in field, at least for the structures
studied here, is of secondary importance for emission wavelength and
radiative lifetime of $c$-plane dots with In contents in the range
of 10\% to 20\%. However, for In contents of order 40\%, our
calculations show that second-order piezoelectricity has a
significant effect on these quantities. For instance, at $x=0.4$, we
observe that for the chosen QD geometry the radiative lifetime is
more than doubled when comparing a calculation that includes
second-order piezoelectric effects to one neglecting this
contribution. Consequently, second-order piezoelectricity has to be
taken into account when designing nanostructures operating in the
long wavelength, high In content regime.

The total strain induced piezoelectric polarization
($\mathbf{P}^\text{Tot}_\text{pz}$) in a semiconductor material with
a lack of inversion symmetry  can be written, up to second-order,
as~\cite{BeZu2006}
\begin{equation}
P^\text{Tot}_{\text{pz},\mu} =
P^\text{FO}_{\text{pz},\mu}+P^{\text{SO}}_{\text{pz},\mu}=\sum_{j=1}^6
e_{\mu j}\epsilon_j + \frac{1}{2}\sum_{j,k=1}^6 B_{\mu
jk}\epsilon_j\epsilon_k\, . \label{eq:piezo_general}
\end{equation}
Here $P^\text{FO}_{\text{pz},\mu} =\sum_{j=1}^6 e_{\mu j}\epsilon_j$
is the first order contribution and $P^\text{SO}_{\text{pz},\mu} =
\frac{1}{2}\sum_{jk=1}^6 B_{\mu jk}\epsilon_j\epsilon_k$ is the
second-order part. The first-order piezoelectric coefficients are
denoted by $e_{\mu j}$ and $B_{\mu jk}$ are second-order ones. The
strain tensor components (in Voigt notation) are given by
$\epsilon_j$. From Eq.~(\ref{eq:piezo_general}) one can infer
already that second-order piezoelectricity should become important
for systems under large strains, in our case high In contents, given
that it is related to products of strain tensor components.

For wurtzite semiconductors the well known first-order contribution
$\mathbf{P}^\text{FO}_\text{pz}$ has  only three independent
piezoelectric coefficients, namely $e_{33}$, $e_{15}$ and
$e_{31}$.~\cite{AnOr2000} For the second-order coefficients $B_{\mu
jk}$, Grimmer~\cite{Gr2007} showed that out of 36 $B_{\mu jk}$
coefficients, 17 are nonzero of which 8 are independent. Taking all
this into account and using cartesian notation for the strain
tensor, in a wurtzite $c$-plane system the total (first plus
second-order) piezoelectric polarization vector field
$\mathbf{P}^\text{Tot}_\text{pz}$ is given by

\begin{widetext}
\begin{eqnarray}
\nonumber \mathbf{P}^\text{Tot}_\text{pz}&=&\begin{pmatrix}
2e_{15}\epsilon_{xz}\\
2e_{15}\epsilon_{yz}\\
e_{31}(\epsilon_{xx}+\epsilon_{yy})+e_{33}\epsilon_{zz}
\end{pmatrix}\\
&& +
\begin{pmatrix}
2B_{115}(\epsilon_{xx}\epsilon_{xz}+\epsilon_{xy}\epsilon_{yz})+2B_{135}\epsilon_{zz}\epsilon_{xz}-2B_{125}(\epsilon_{xy}\epsilon_{yz}-\epsilon_{yy}\epsilon_{xz})\\
2B_{115}(\epsilon_{yy}\epsilon_{yz}+\epsilon_{xy}\epsilon_{xz})+2B_{135}\epsilon_{zz}\epsilon_{yz}+2B_{125}(\epsilon_{xx}\epsilon_{yz}-\epsilon_{xy}\epsilon_{xz})\\
\frac{B_{311}}{2}(\epsilon_{xx}^2+\epsilon_{yy}^2+2\epsilon_{xy}^2)+B_{312}(\epsilon_{xx}\epsilon_{yy}-\epsilon_{xy}^2)+
B_{313}(\epsilon_{xx}\epsilon_{zz}+\epsilon_{yy}\epsilon_{zz})+2B_{344}(\epsilon_{yz}^2+\epsilon_{xz}^2)+\frac{B_{333}}{2}\epsilon_{zz}^2
\end{pmatrix}.
\label{eq:Full_Vector}
\end{eqnarray}
\end{widetext}

It should be noted that this expression is far more complicated when
compared to piezoelectric effects in zincblende structures. In the
zincblende case one is left with only one first-order and three
independent second-order piezoelectric
coefficients.~\cite{BeZu2006,Gr2007}

\begin{figure*}[t!]
\begin{tabular}{|c|c|c|}
\hline \textbf{(a) First-order + spontaneous} & \textbf{(b)
Second-order} & \textbf{(c) Total}\\\hline
  \includegraphics[width=0.3\textwidth]{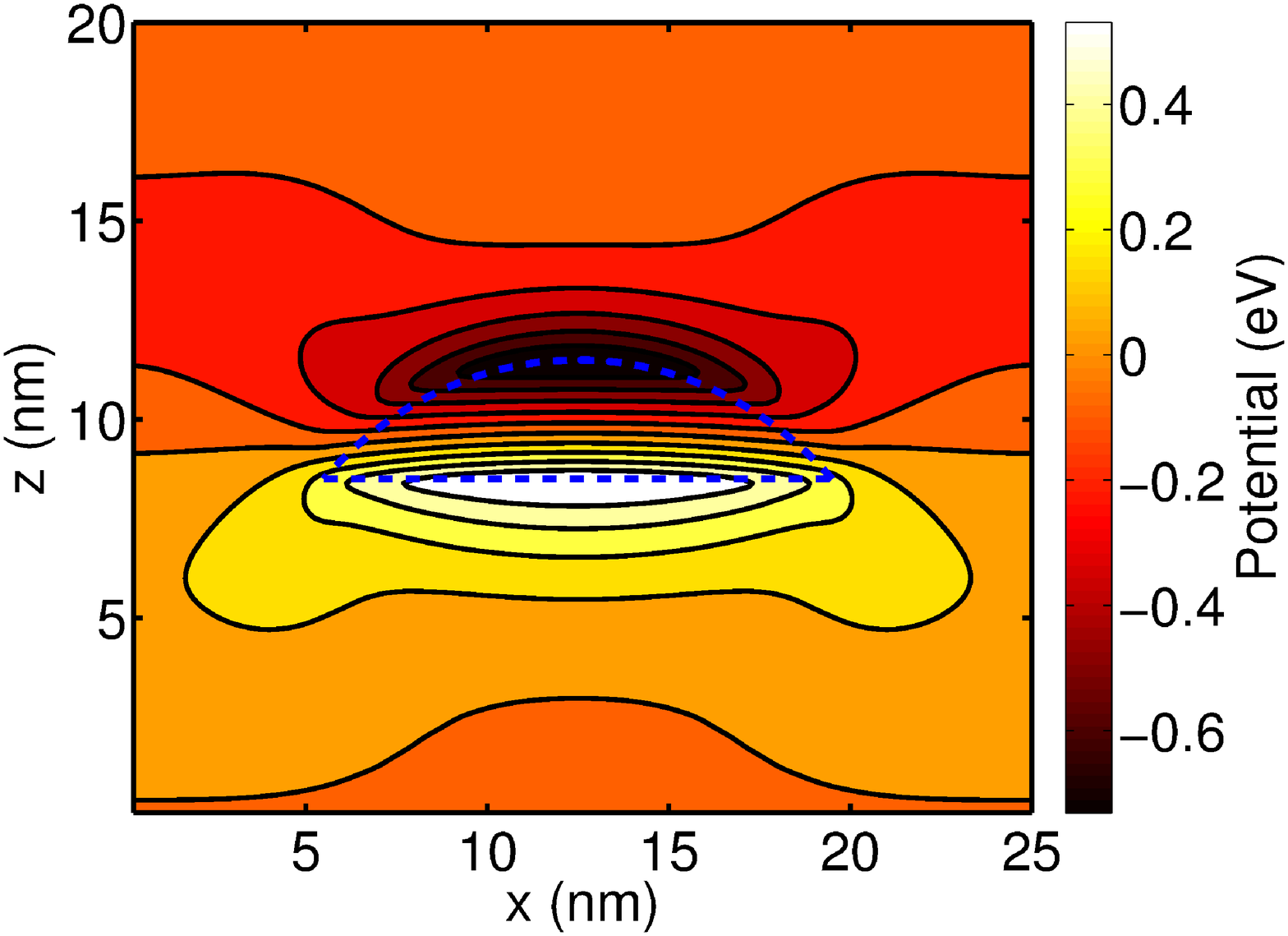} &
  \includegraphics[width=0.3\textwidth]{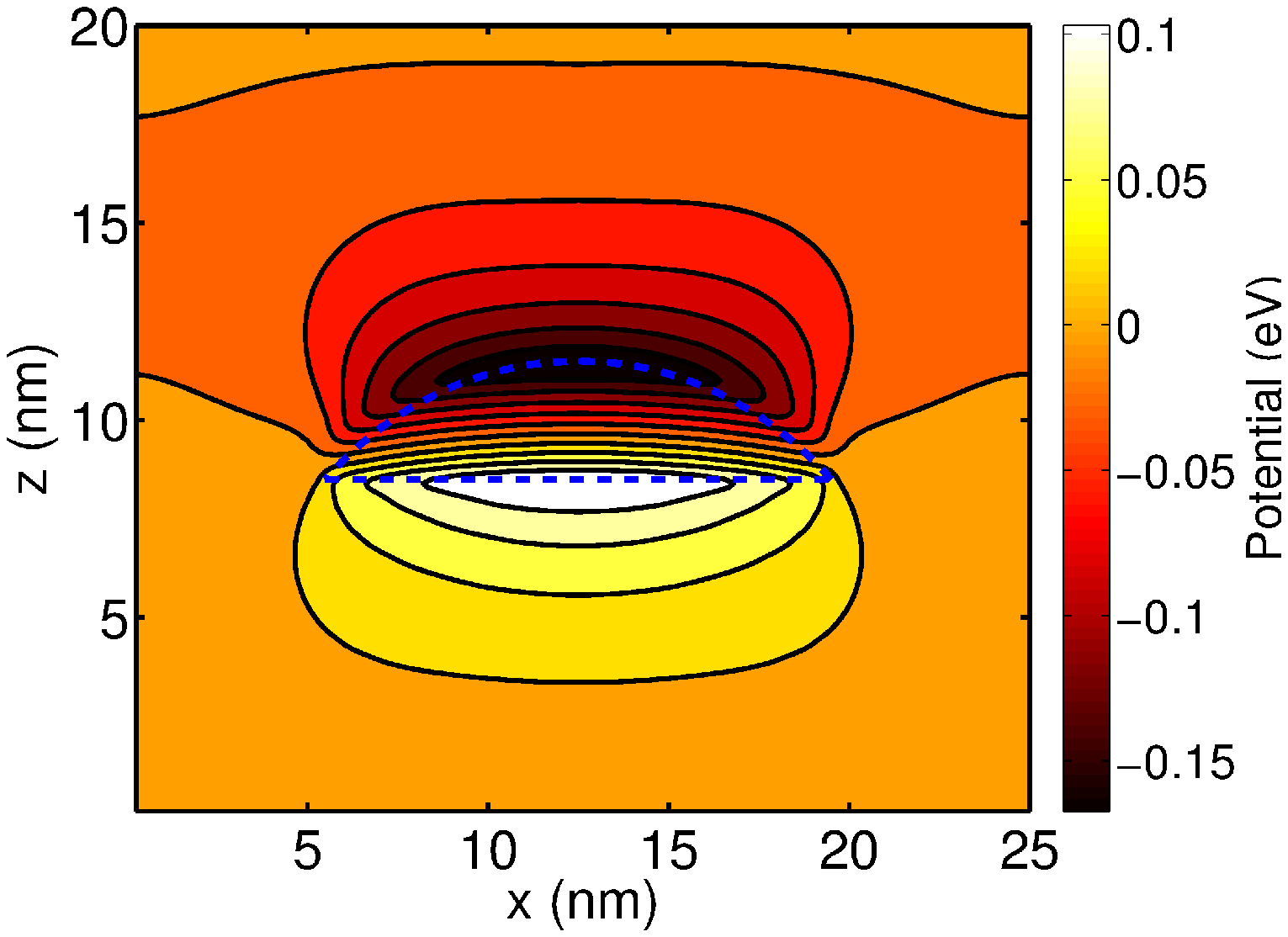} &
  \includegraphics[width=0.3\textwidth]{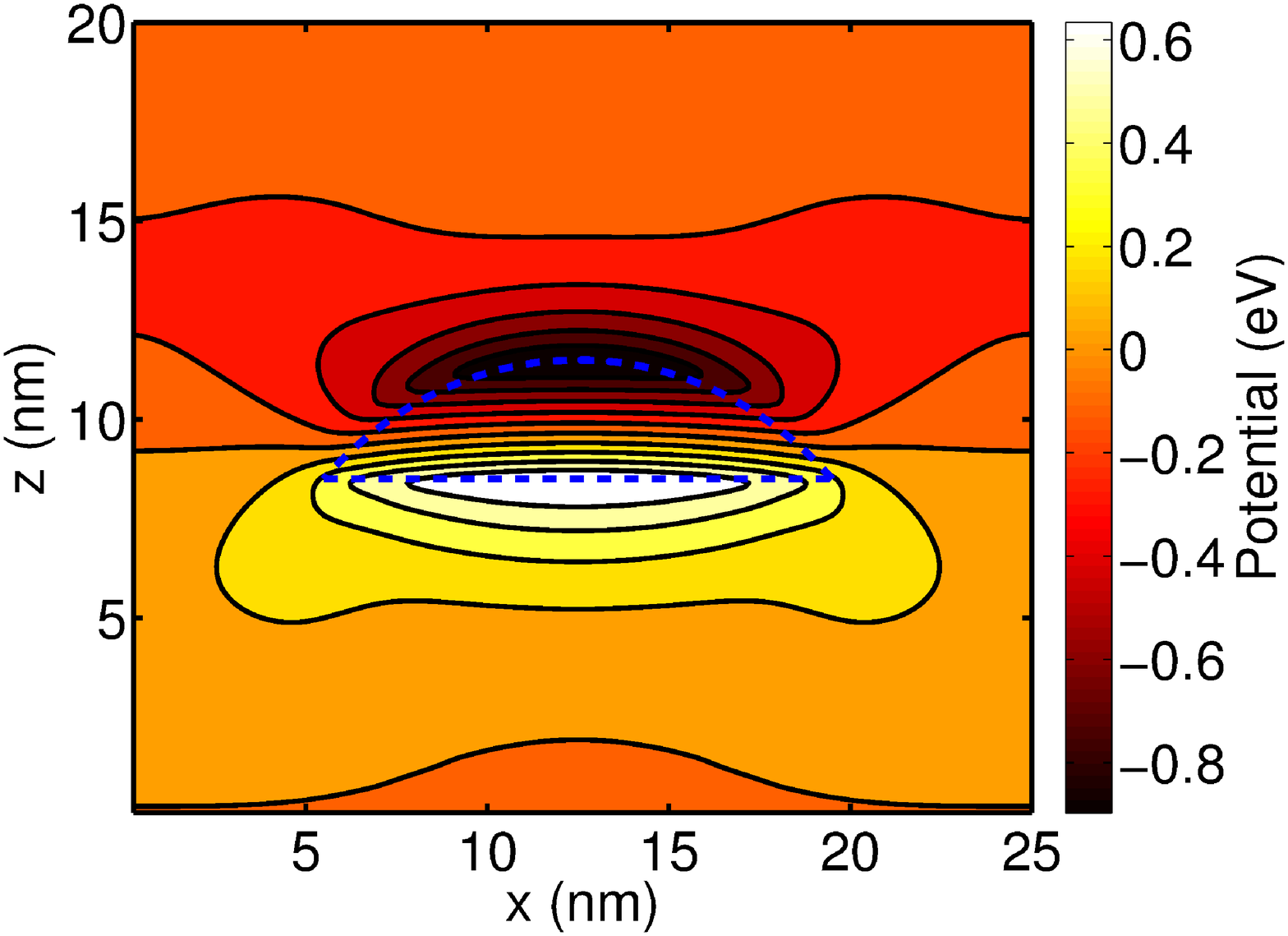}\\\hline
\end{tabular}
\caption{(a) Contour plot of the built-in potential arising from
spontaneous and first-order piezoelectric polarization in a
lens-shaped In$_{0.5}$Ga$_{0.5}$N/GaN $c$-plane dot with a base
diameter $d=14$ nm and a height $h=3$ nm. The contour plot is shown
for a slice through the center of the dot in the $x-z$ plane. (b)
Same as in (a) but here second-order piezoelectric effects only are
considered. (c) Same as in (a) but the total
(spontaneous+first-order+second-order) built-in potential is shown.}
\label{fig:potential}
\end{figure*}

To analyze the impact of second-order piezoelectricity on the
electronic and optical properties of $c$-plane
In$_{x}$Ga$_{1-x}$N/GaN QDs with varying In content $x$, we have
performed continuum-based calculations by means of
$\mathbf{k}\cdot\mathbf{p}$ theory. All calculations have been
carried out in the highly flexible plane wave based software library
S/PHI/nX,~\cite{MaBo2014} allowing us to define customized
piezoelectric polarization vector fields such as
Eq.~(\ref{eq:Full_Vector}). In doing so, we are able to perform
calculations in the presence and absence of second-order
piezoelectricity. Additionally, spontaneous polarization effects are
included in the calculations, with values for GaN and InN taken from
Ref.~\onlinecite{CaSc2013}, and the bowing parameter from
Ref.~\onlinecite{VuMe2013}. First-order piezoelectric coefficients
from Ref.~\onlinecite{CaSc2013} and second-order coefficients from
Ref.~\onlinecite{PrWa2013} have been used. For the electronic
structure calculations we have applied a six-band Hamiltonian to
describe the hole states and a single-band effective mass
approximation for electron states. This model accounts for valence
band mixing effects and the differences in the effective masses
along different directions. Since we are mainly interested in e.g.
wavelength shifts due to second-order piezoelectricity, the applied
electronic structure approach is sufficient for our purposes.
Further refinements can be achieved by applying an eight-band
model.~\cite{MiYa2012} Using the $\mathbf{k}\cdot\mathbf{p}$ wave
functions, the radiative lifetime $\tau$ of the electron and hole
ground state transition has been
calculated,~\cite{PaWa2017,FoBa2003} employing a light polarization
vector $\mathbf{e}$ perpendicular to the sample surface. More
details on theoretical framework and parameter sets applied are
given in Refs.~\onlinecite{PaSc2017,MaBo2014,PaMa2016,PaWa2017}.

Following previous studies on In$_{x}$Ga$_{1-x}$N QDs, we have
assumed a lens-shaped dot geometry.~\cite{BaSc2013,ScRe2010} Based
on earlier atomic force microscopy results, a QD base diameter of
$d=14$ nm and a dot height of $h=3$ nm has been
chosen.~\cite{MoSo2000,SeSm2007} Here our main focus is on how
electronic and optical properties of $c$-plane In$_{x}$Ga$_{1-x}$N
QDs change with increasing In content $x$ when second-order
piezoelectric effects are considered. Thus, we vary the In content
$x$ of the dot between 10\% and 50\% in 10\% steps.

\begin{figure}
\begin{tabular}{|c|}
\hline
\textbf{First-order + spontaneous}\\
\includegraphics[width=0.725\columnwidth]{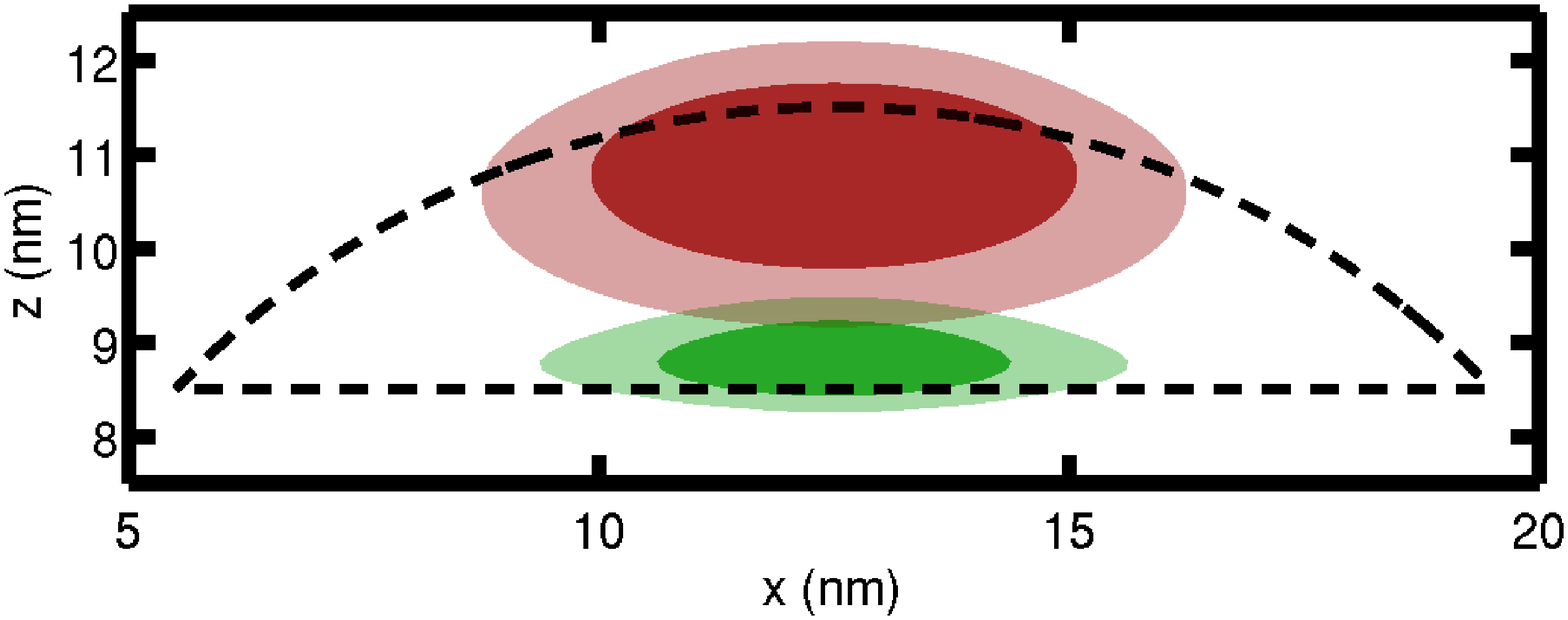}\\\hline
\textbf{Total}\\
\includegraphics[width=0.725\columnwidth]{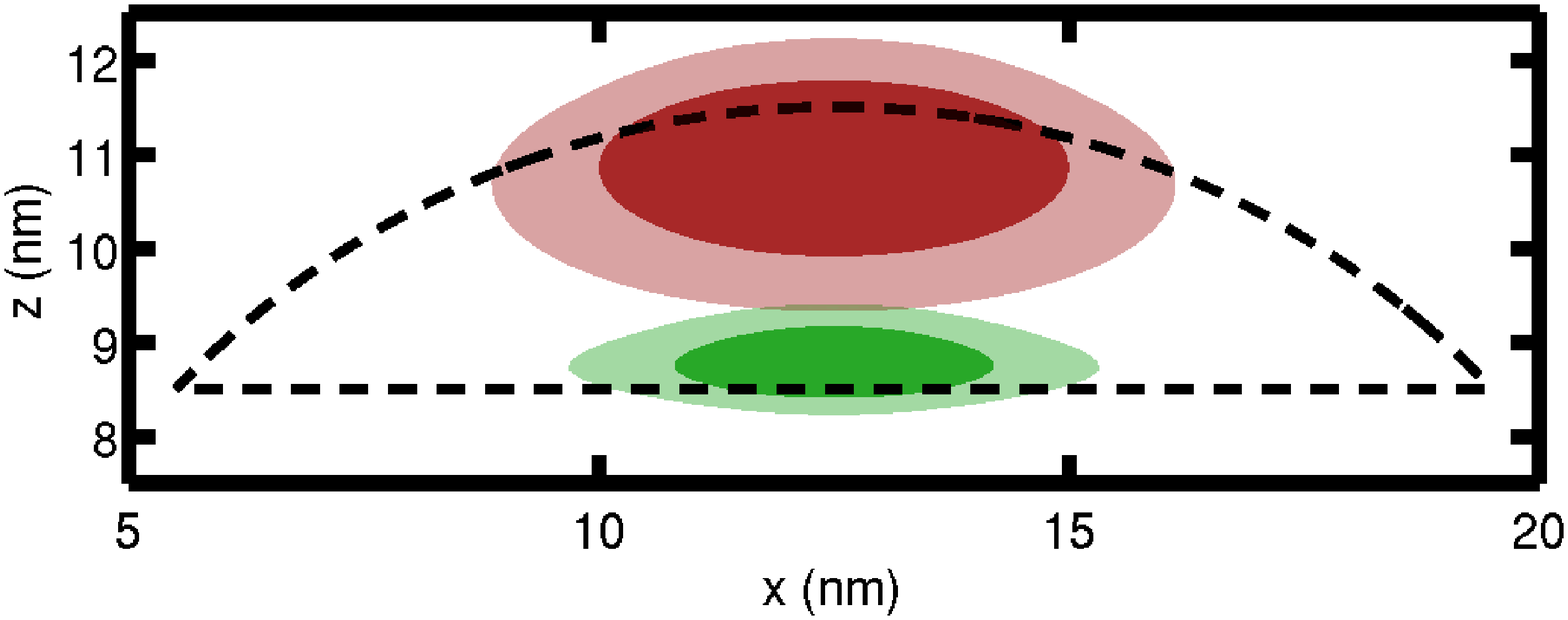}\\\hline
\end{tabular}
\caption{Isosurface plots of the electron (red) and hole (green)
ground state charge densities at 5\% (light surface) and 25\% (dark
surface). The QD geometry is indicated by the dashed line. (a)
Built-in potential due to spontaneous and first-order piezoelectric
polarization only. (b) Full built-in potential.}
\label{fig:wavefunctions}
\end{figure}

In a first step we analyze how the built-in potential in a $c$-plane
In$_{x}$Ga$_{1-x}$N/GaN QD is changed when including second-order
piezoelectric effects. Contour plots of the built-in potential of a
$c$-plane In$_{0.5}$Ga$_{0.5}$N dot, for a slice through the QD
center, are shown in Fig.~\ref{fig:potential}. The slice is taken in
the $x-z$-plane, where the $z$-axis is parallel to the wurtzite
$c$-axis. The impact of changes in the In content $x$ will be
investigated below when we discuss electronic and optical properties
of the structures under consideration. In Fig.~\ref{fig:potential}
(a) the built-in potential arising from the standard first-order
piezoelectric contribution and the spontaneous polarization is
shown. The well known potential drop along the $c$-axis, leading to
a spatial separation of electron and hole wave functions, is clearly
visible. Figure~\ref{fig:potential} (b) depicts the second-order
piezoelectric contribution only. Here, several features are of
interest. First, the magnitude of the second-order contribution,
even at $x=0.5$, is a factor of order 4 smaller compared to the
situation where only first-order piezoelectricity and spontaneous
polarization are taken into account (cf. Fig.~\ref{fig:potential}
(a)). Nevertheless, the second-order contribution has still a
sizeable magnitude. Second, the symmetry of the second-order
contribution and the potential profile are similar to the
first-order contribution (cf. Fig.~\ref{fig:potential} (a)). Thus,
when taking second-order piezoelectric effects into account, the
potential drop across the QD will be larger, leading  to an even
stronger electrostatic built-in field and thus to an even stronger
spatial separation of electron and hole wave functions. We will come
back to this effect below. The fact that the second-order built-in
potential contribution is of the same symmetry as the first-order
term is for instance different to zincblende InAs/GaAs
QDs.~\cite{BeZu2006} Additionally, it should be noted that the
magnitude of the second-order piezoelectric contribution also
depends on the QD shape and size, as highlighted by Schliwa \emph{et
al.}~\cite{ScWi2007} for InGaAs/GaAs QD systems. To gain initial
insights into this question for InGaN QDs, we have performed
additional calculations for a slightly larger dot ($d=18$ nm, $h=3$
nm) with 30\% In. This study reveals only a slight increase in the
potential drop across the nanostructure when compared to a
In$_{0.3}$Ga$_{0.7}$N dot with $d=14$ nm and $h=3$ nm. For different
QD geometries and/or higher dots this situation might change.
However, a detailed analysis of the impact of QD shape and size is
beyond the scope of the present study. Here, we are interested in
establishing trends with increasing In content.
Figure~\ref{fig:potential} (c) shows the total
(spontaneous+first-order+second-order) built-in potential for the
considered lens-shaped $c$-plane In$_{0.5}$Ga$_{0.5}$N QD. As
expected from the discussion above, when including second-order
piezoelectric effects, the total potential drop is clearly increased
compared to the situation where only first-order piezoelectricity
and spontaneous polarization are considered (cf.
Fig.~\ref{fig:potential} (a)). The question is now, how strongly are
electronic and optical properties affected when taking second-order
piezoelectric effects into account? In the following we will look at
the impact of second-order piezoelectricity on the emission
wavelength $\lambda$ and the radiative lifetime $\tau$ as a function
of the dot In content $x$. But before turning to these questions, we
start with looking at the electron and hole ground state charge
densities of the In$_{0.5}$Ga$_{0.5}$N QD.
Figure~\ref{fig:wavefunctions} shows the isosurfaces of the electron
(red) and hole (green) ground state charge densities. In
Fig.~\ref{fig:wavefunctions} (a) results in the absence of the
second-order piezoelectric effect are shown; (b) depicts the data
originating from a calculation accounting for the full built-in
potential, thus including second-order piezoelectricity. From
Fig.~\ref{fig:wavefunctions} we can conclude that the increase in
the built-in potential due to second-order piezoelectricity leads to
a stronger spatial separation of the carriers along the $c$-axis.
Thus the wave function overlap is reduced and consequently the
radiative lifetime will increase. Also, due to the increased
built-in potential, an increased red shift of the emission
wavelength due to second-order piezoelectricity is expected.

\begin{figure}[t]
\includegraphics[width=0.725\columnwidth]{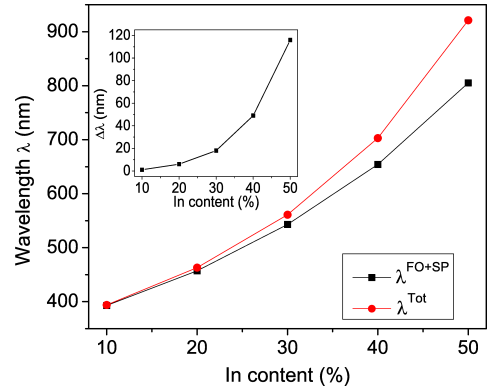}
\caption{\label{fig:wavelength} Emission wavelength $\lambda$ as a
function of In content $x$. Results in absence of second-order
piezoelectricity, taking only spontaneous (SP) and first-order (FO)
piezoelectric polarization into account, are given by the black
squares ($\lambda^\text{FO+SP}$). The red circles denote data when
including second-order piezoelectricity ($\lambda^\text{Tot}$). The
inset shows
$\Delta\lambda=\lambda^\text{Tot}-\lambda^\text{FO+SP}$.}
\end{figure}

Figure~\ref{fig:wavelength} shows the emission wavelength $\lambda$
of the here considered $c$-plane In$_{x}$Ga$_{1-x}$N/GaN QD for In
contents varying between 10\% and 50\%. The black squares
($\lambda^\text{FO+SP}$) denote the results in the absence of
second-order contributions [only first-order (FO) piezoelectricity
and spontaneous (SP) polarization], while the red circles
($\lambda^\text{Tot}$) show the data when including also
second-order piezoelectric effects. Again, it should be noted that
for emitters operating in the red wavelength regime, In contents of
40\% have been reported in the literature,~\cite{FrHa2016} so that
the here studied In content range is relevant to recent experimental
studies. From Fig.~\ref{fig:wavelength} one can infer that for lower
In contents (up to 20\%), the second-order piezoelectric
contribution has little effect on $\lambda$. In fact in this case
the difference in the emission wavelength $\Delta
\lambda=\lambda^\text{Tot}-\lambda^\text{FO+SP}$, obtained from a
calculation with spontaneous and first-order piezoelectric
polarization only, $\lambda^\text{FO+SP}$, and a calculation
including second-order piezoelectric effects, $\lambda^\text{Tot}$,
is less than 10 nm. To show this effect more clearly, the inset in
Fig.~\ref{fig:wavelength} depicts $\Delta \lambda$ as a function of
the In content $x$. Between 30\% and 40\% In, second-order effects
lead to a noticeable difference, resulting in $\Delta \lambda$
values of approximately 20 nm to 50 nm, respectively. At 50\% In we
observe a wavelength shift of $\Delta \lambda=120$ nm. Overall, the
wavelength shift is almost equally distributed between electron and
hole ground state energy shifts. We attribute this to the combined
effect of differences in electron and hole effective masses and the
asymmetry in the magnitude of the built-in potential between the
upper and lower QD interface. Moreover, the change in the
confinement potential due to second-order piezoelectricity might
also affect the Coulomb interaction between the carriers and can
lead to further contributions to the wavelength shift discussed here
in the single-particle picture. Overall, our calculations reveal two
things. First, when targeting QD-based emitters operating in the red
spectral regime ($\approx$ 650 nm), second-order piezoelectric
effects can play a significant role. Furthermore, the second-order
piezoelectric contribution shifts the emission to longer wavelength.
Thus, when designing emitters operating in this long wavelength
regime, the required In content predicted from a model including
second-order effects would be lower as expected from a ``standard
model'', which accounts for first-order piezoelectric effects and
spontaneous polarization only.

\begin{figure}[t!]
\includegraphics[width=0.725\columnwidth]{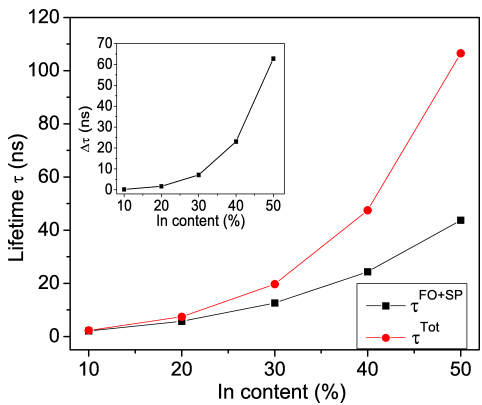}
\caption{\label{fig:radiative} Radiative lifetime $\tau$ as a
function of In content $x$. Results in absence of second-order
piezoelectricity, including spontaneous (SP) and first-order (FO)
piezoelectric polarization only, are given by the black squares
($\tau^\text{FO+SP}$). The red circles denote the data when
including second-order piezoelectricity ($\tau^\text{Tot}$). The
inset shows $\Delta \tau=\tau^\text{Tot}-\tau^\text{FO+SP}$.}
\end{figure}

Even though our analysis indicates that lower In contents are
sufficient to reach emission at longer wavelength, the increase in
the built-in potential responsible for this effect will have a
detrimental effect on the wave function overlap and consequently on
the radiative lifetime $\tau$. To study the impact of second-order
piezoelectricity on the radiative lifetime $\tau$,
Fig.~\ref{fig:radiative} depicts $\tau$ in the absence
($\tau^\text{FO+SP}$, black squares) and in the presence
($\tau^\text{Tot}$, red circles) of second-order piezoelectric
contributions. Similar to the wavelength shift discussed above, in
the In content range of 10\% to 20\% the influence of second-order
piezoelectricity is of secondary importance ($\Delta\tau\leq 2$ ns).
The inset of Fig.~\ref{fig:radiative} depicts the difference in the
radiative lifetime $\Delta \tau=\tau^\text{Tot}-\tau^\text{FO+SP}$,
obtained from calculations including ($\tau^\text{Tot}$) and
neglecting ($\tau^\text{FO+SP}$) second-order piezoelectric
contributions. The calculated radiative lifetimes in the 10\% to
20\% In regime are in the range of 3 ns to 10 ns, which is in good
agreement with reported experimental data on these
systems~\cite{RoRi2003,JaOl2007}. However, for higher In contents we
clearly observe a significant contribution from second-order
piezoelectricity. At 30\% the $\tau$ value is a factor of order 1.5
larger ($\tau^\text{FO+SP}=12$ ns; $\tau^\text{Tot}=19$ ns) when
including second-order piezoelectric effects in the calculations. At
40\% and 50\% In, the value of $\Delta \tau$ becomes 23 ns and 62
ns, respectively. But, it should be noted that the here calculated
radiative lifetimes for a $c$-plane In$_{0.4}$Ga$_{0.6}$N QD, even
without second-order effects, are much larger than the experimental
values ($\tau^\text{exp}=3$ ns) reported in the literature for InGaN
dots with 40\% In.~\cite{FrHa2016} Further studies, both
theoretically and experimentally, are required to shed more light
onto the physics of In$_{x}$Ga$_{1-x}$N QDs operating in the long
wavelengths regime (green to red).

In summary, we have presented a detailed analysis of the impact of
second-order piezoelectricity on the electronic and optical
properties of $c$-plane In$_{x}$Ga$_{1-x}$N/GaN QDs. Our study
revealed that the second-order piezoelectric effect leads to an
increase in the built-in field when compared to calculations taking
only first-order piezoelectricity and spontaneous polarization into
account. However, when looking at emission wavelength shifts or
radiative lifetime values, at In contents around 10\% to 20\%, these
quantities are almost unaffected by second-order piezoelectricity.
But, when exceeding 30\% In, both quantities are affected
significantly by second-order contributions. The second-order
piezoelectric effect induced built-in field increase leads to the
situation that the emission is shifted to longer wavelength in
comparison to a calculation based on spontaneous and first-order
piezoelectric polarization effects only. This means that when
accounting for second-order effects, lower In contents can be
considered to reach for instance emission in the red spectral
region. On the other hand, the increase in the built-in potential
due to second-order piezoelectric contributions results in a strong
increase in the radiative lifetime for long wavelength, high In
content emitters when compared to results from a ``standard''
first-order study. Overall, our results reveal that when targeting
In$_{x}$Ga$_{1-x}$N QD-based emitters operating in the yellow to red
spectral regime, second-order piezoelectricity cannot be neglected
and should be taken into account for designing and understanding the
electronic and optical properties of these systems.

This  work  was  supported  by  Science  Foundation  Ireland
(project  number  13/SIRG/2210). The authors would like to thank
Brian Corbett, Eoin P. O'Reilly and Miguel A. Caro for fruitful
discussions.

%

\end{document}